\documentclass[conference]{IEEEtran}
\IEEEoverridecommandlockouts
\usepackage{cite}
\usepackage{amsmath,amssymb,amsfonts}
\usepackage{algorithmic}
\usepackage{graphicx}
\usepackage{textcomp}
\usepackage{xcolor}
\usepackage{balance} 
\usepackage{booktabs}
 \usepackage{subcaption}
\usepackage{comment}
\def\BibTeX{{\rm B\kern-.05em{\sc i\kern-.025em b}\kern-.08em
    T\kern-.1667em\lower.7ex\hbox{E}\kern-.125emX}}

\begin{document}
\pagenumbering{roman}
\title{Cognitive Engagement for STEM+C Education: Investigating Serious Game Impact on Graph Structure Learning with fNIRS}

\author{\IEEEauthorblockN{Shayla Sharmin$^{1}$, 
Reza Koiler$^{2}$,  
Rifat Sadik$^{1}$, 
Arpan Bhattacharjee$^{1}$, 
Priyanka Raju Patre$^{1}$, 
\\ Pinar Kullu$^{1}$,
Charles Hohensee$^{3}$, 
Nancy Getchell$^{2}$, 
Roghayeh Leila Barmaki$^{1}$ 
}
\IEEEauthorblockA{$^{1}$Department of Computer and Information Sciences, University of Delaware, USA
}
\IEEEauthorblockA{$^{2}$Department of Kinesiology \& Applied Physiology, University of Delaware, USA
}
\IEEEauthorblockA{$^{3}$School of Education, University of Delaware, USA
}
}
\maketitle

\begin{abstract}
For serious games on education, understanding the effectiveness of different learning methods in influencing cognitive processes remains a significant challenge. In particular, limited research addresses the comparative effectiveness of serious games and videos in analyzing brain behavior for graph structure learning, which is an important part of the Science, Technology, Engineering, Math, and Computing (STEM+C) disciplinary education.
This study investigates the impact of serious games on graph structure learning. For this, we compared our in-house game-based learning (GBL) and video-based learning (VBL) methodologies by evaluating their effectiveness on cognitive processes by oxygenated hemoglobin levels using functional near-infrared spectroscopy (fNIRS).
We conducted a $2\times 1$ between-subjects preliminary study with twelve participants, involving two conditions: game and video. Both groups received equivalent content related to the basic structure of a graph, with comparable session lengths. The game group interacted with a quiz-based game, while the video group watched a pre-recorded video. The fNIRS was employed to capture cerebral signals from the prefrontal cortex, and participants completed pre- and post-questionnaires capturing user experience and knowledge gain. 
In our study, we noted that the mean levels of oxygenated hemoglobin ($\Delta HbO$) were higher in the GBL group, suggesting the potential enhanced cognitive involvement. Our results show that the lateral prefrontal cortex (LPFC) has greater hemodynamic activity during the learning period. Moreover, knowledge gain analysis showed an increase in mean score in the GBL group compared to the VBL group. Although we did not observe statistically significant changes due to participant variability and sample size, this preliminary work contributes to understanding how GBL and VBL impact cognitive processes, providing insights for enhanced instructional design and educational game development. Additionally, it emphasizes the necessity for further investigation into the impact of GBL on cognitive engagement and learning outcomes.
\end{abstract}

\begin{IEEEkeywords}
serious game, game-based learning, video-based learning, brain activity, fNIRS, hemodynamic response, oxygenated ($\Delta HbO$), and deoxygenated hemoglobin ($\Delta HbR$).
\end{IEEEkeywords}
\section{Introduction} \label{Introduction}
In educational research, a persistent challenge revolves around understanding the efficacy of different learning approaches in influencing cognitive processes and improving learning outcomes \cite{shu2023empirical,ronghuai2014three}. Today’s students spend a significant amount of time engaging with technologies such as mobile devices, laptops, and tablets to watch videos and play games for recreation and education purposes \cite{haleem2022understanding,panjeti2023impact}. As a result, researchers and educators are actively seeking ways to enhance the interactivity and engagement of video and game content. It has been found that motivation and engagement in learning are key factors influencing academic achievement \cite{collie2019motivation} and that results in higher learning outcomes and performance \cite{chattopadhyay2021motivation}. 
The use of video as learning material is very promising in education. Serious games are also considered an effective tool for educational and training purposes \cite{ninaus2014neurophysiological,desoto2023utilization,samah2018using}. Thus, the effectiveness of game-based learning (GBL) and video-based learning (VBL) in enhancing cognitive processes has received significant attention. However, investigating this efficacy has often relied on self-reported outcomes from observational studies, leaving a gap in understanding the underlying neural processing. 

Analyzing neural activity can help identify the key insights into how the brain responds to various learning techniques that keep students cognitively engaged \cite{materna2007jump}. Despite advancements in methodologies for investigating the interconnected interaction between the brain and behavior, there is a lack of studies dedicated to the development of integrated models or explanatory frameworks that encompass both neuroscience, which focuses on the study of the brain, and cognition, which pertains to the processes underlying human thinking and behavior \cite{ghaderi2023general}. Measuring brain activity allows researchers and educators to gain insights into how the brain processes information, what parts of the brain are active during different learning tasks, and how different learning strategies affect brain function. However, it is important to note that brain signal analysis is still a growing research field because cognitive processes cannot be observed directly. Instead, they are inferred indirectly from factors such as task performance and hemodynamic responses \cite{heathcote2015introduction}. Furthermore, cognitive abilities vary from individual to individual due to different neural processing speeds \cite{schubert2019individual}. So, more research is needed to fully understand how best to use these techniques for educational purposes. 

Research in the prefrontal cortex has made invaluable progress in brain signal analysis, as it involves many higher cognitive functions, such as decision-making, working memory, and attention \cite{kober2020game,doherty2023interdisciplinary}. Functional Near-Infrared Spectroscopy (fNIRS) can be used to determine neural activity in the brain from the prefrontal cortex \cite{lloyd2010illuminating, skau2021exhaustion}. It is a non-invasive optical neuroimaging technique used to measure changes in optical density that are correlated with hemodynamic response and neural activity \cite{ninaus2014neurophysiological}. fNIRS calculate hemodynamic responses by measuring $\Delta HbO$ which refers to changes in oxygenated hemoglobin (HbO) concentration and $\Delta HbR$ deoxygenated hemoglobin (HbR) in the prefrontal cortex \cite{doherty2023interdisciplinary,li2023current}.
Cortical activity results in an inflow of oxygenated blood \cite{koiler2022impact}. The active neuron consumes oxygen when the level of neural activity in a brain area initially increases $\Delta HbR$ \cite{koiler2022impact,shealy2023changes,cakir2015optical}. When the demand for oxygen increases, oxygen flows to this area of the brain, increasing levels of $\Delta HbO$ \cite{koiler2022impact,cakir2015optical}. 
 Higher $\Delta HbO$ and $\Delta HbR$ indicate more blood flow to the cortex and, therefore, stronger brain activities in neurons while performing a task \cite{koiler2022impact,cakir2015optical,ccakir2016behavioral}.
However, most studies investigate brain activity for a specific game \cite{kober2020game,cakir2015optical,ccakir2016behavioral,greipl2021brain,samah2018using} or video \cite{desoto2023utilization,tang2023mind} interface. To our knowledge, no study has compared the effectiveness of both learning techniques. The comparative analysis becomes more valid if we can study brain activity for video and game-based interfaces for a similar topic. The application of fNIRS to investigate the differences in cortical activity associated with forms of learning can be used to elucidate the underlying neural mechanisms and neural bio-markers of learning associated with each method. Therefore, analyzing and comparing the brain signals of learners playing an educational game and watching videos will be important in cognitive neuroscience. It can also help educators to design study materials and make the learning process more interactive.
This study aims to bridge this gap by conducting a comparative analysis of GBL and VBL using brain behavior analysis, specifically measuring oxygenated hemoglobin using fNIRS. The goal of this study is to explore the interplay between learning methodologies and cognitive processes, providing valuable insights for designing and implementing more effective educational strategies. We seek to illuminate the extent to which GBL and VBL impact cognitive engagement and knowledge acquisition. The results of this study have the potential to enhance the field of education and human-computer interaction by providing evidence-based guidance for designing immersive and impactful learning experiences.
Our investigation focuses on brain activity, changes in engagement, and knowledge acquisition while learning computer science topics, such as the basic structure of graphs, in a between-subject pilot study. Engagement is assessed by changes in fNIRS measurements of $\Delta HbO$ and $\Delta HbR$, while knowledge gain is evaluated by comparing pre- and post-test scores.
\begin{itemize}
    \item \textbf{RQ1:} What are the differences in neural activities between game-based and video-based learning?
    \vspace{.7em}
    \item \textbf{RQ2:} What are the differences regarding the subjective results on usability, task load, and knowledge gain between game-based and video-based learning modules? 
\end{itemize}

\vspace{1em}

In this study, our contribution is to compare GBL vs VBL within the context of graph theory by measuring the hemodynamic response. While previous research has primarily focused on assessing user engagement and knowledge scores to evaluate these two learning approaches, our work uses fNIRS data as a way to provide a more comprehensive understanding. Our aim was to investigate whether traditional measures of learning effectiveness, such as user engagement and knowledge acquisition, are supported by physiological evidence, particularly regarding brain activity as indicated by the fNIRS data. Given that earlier studies have often found GBL to be more effective, as seen through better performance scores, our study sought to determine whether these scores are associated with the fNIRS data. Our approach combines subjective learning effectiveness indicators with objective brain physiological data to provide a more holistic perspective of how different educational modalities affect learning.

The remainder of this paper is organized as follows. Section \ref{Related Works} presents related works on game- and video-based learning and analysis of brain signals. Section \ref{Materials and Methods} describes the proposed methodology and study design. Section \ref{Result} discusses the results of our experiments.  Finally, Section \ref{Conclusion} concludes the paper by discussing study findings, the proposed system's limitations, and the scope for future work.

\section{Related Work}\label{Related Works}


Analyzing brain signals is of importance, as it enables researchers to examine the neural underpinnings of cognitive processes and the mental workload of the human cortex \cite{shealy2023changes,essa2021brain}. Functional Near Infrared Spectroscopy (fNIRS) \cite{ayaz2012optical},  electroencephalogram (EEG) \cite{samah2018using}, and functional Magnetic Resonance Imaging (fMRI) \cite{greipl2021brain} are scientific methodologies that offer significant contributions to the understanding of brain activities. These techniques are particularly useful in evaluating attention, engagement, and cognitive load. Functional near-infrared spectroscopy, as a brain imaging technique that is comparatively more cost-effective, lightweight, and portable, plays a significant role in facilitating the analysis of brain signals flexibly \cite{doherty2023interdisciplinary,koiler2022impact,wei2023reduced,wang2023interaction}.


 By studying brain signals, educators can design learning materials that align with cognitive processes, optimize learning experiences, and improve retention and understanding \cite{chang2021neuroscience}. Brain signal analysis techniques fNIRS and EEG have been used in STEM including math \cite{cakir2015optical,skau2022proactive,poikonen2023nonlinear}, geometry \cite{shi2023improving}, engineering \cite{shealy2023changes,grohs2017evaluating}, and science \cite{naimi2010investigating} to investigate neural activity and cognitive processes. 

 \subsection{Learning and Performance: Cognitive and Neural Factors}
Proactive control, hands-on learning, thinking aloud, and mathematical performance are reviewed in this section. Besides, cognitive processes, brain activation, and mathematical proficiency are discussed.
In a study, Suko et al. \cite{skau2022proactive} examined mathematical cognition, general cognition, and the brain bases in 8- to 9-year-olds. The study showed that proactive control correlates more strongly with mathematical performance than with other cognitive abilities using additive mathematics tests, cognitive assessments, and fNIRS brain imaging.
Shi et al. 
\cite{shi2023improving} observed that hands-on experience led to an increase in the concentration of oxygenated hemoglobin, as measured by fNIRS. This increase indicated higher neuron activation compared to video teaching with 40 Chinese middle school students.
Tripp et al. 
\cite{shealy2023changes} found that the act of expressing thoughts, commonly referred to as thinking aloud, significantly impacts both the cognitive processes involved in design and the neurological processes underlying cognition. They used fNIRS to measure the changes in oxygenated hemoglobin.
Artemenko et al.
\cite{artemenko2019individual} observed a correlation between lower mathematical proficiency and shorter calculation duration, particularly for complex arithmetic problems. Individuals with lower mathematical skills also displayed reduced neural activity in the left supramarginal, superior temporal, and inferior frontal gyri. The study employed both EEG and fNIRS to observe these neural activities.


\subsection{Brain Signal Analysis in Serious Game and Video: Enhancing Neural Activity and Engagement}
Several studies, including that by Desoto et al. \cite{desoto2023utilization}, have explored brain signal analysis during activities like playing games and watching videos. They observed that different inputs can trigger unique neural responses, suggesting that brains may process similar information in varied ways. Techniques like EEG and fNIRS have been used to monitor cerebral activity during the viewing of STEM educational movies, aiming to understand how the brain interprets identical information.

Tang et al. \cite{tang2023mind} identified mind wandering during VBL using EEG and machine learning techniques. The average area under the receiver operating characteristic curve for classifying mind wandering within individuals was 0.876. Across educational lectures, it was 0.703. This indicates that their system is more effective than random guessing at detecting mind wandering but less precise than individual-level detection.
A study by  Cakir et al. 
\cite{cakir2015optical} aimed to assess the efficacy of GBL in enhancing math fluency. Twenty-seven college students participated in this study, where the game group played ``MathDash," and the control group faced a drill and practice approach to evaluate the behavioral and neural effects. The fNIRS device was used to collect brain signals by measuring the oxygenated hemoglobin $\Delta HbO$. The $\Delta HbO$ was equal in both control groups. Still, during the post-test, the game group had a higher $\Delta HbR$ concentration, which was interpreted by the authors to mean that their game training optimizes their brain metabolism. 

Samah et al.
\cite{samah2018using} examined brain functional connectivity during game-based problem-solving tasks. The primary focus was to draw gender differences based on brain signals while playing a game. For this purpose, they chose a computer-based Tower of Hanoi game. EEG signals were collected in this study to record participant performance, and partial directed coherence (PDC) analysis was performed to analyze the data. PDC is a statistical method used to analyze the directional interactions between different time series data, such as EEG signals. PDC analysis illustrates the interactions between time direction and spectral properties of a signal in the brain. According to the study, male and female respondents exhibited no appreciable differences in brain activity patterns.
 In another study using near-infrared spectroscopy, Kober et al.
 \cite{kober2020game} investigated behavioral performance while learning by playing games on a neurofunctional level. They used a NIRS device to calculate the frontal brain's oxygenated hemoglobin and deoxygenated hemoglobin concentration from 59 healthy adults. They found that the game version was more engaging than the non-game version by observing more robust activation in the prefrontal cortex.  They also used the Flow Short Scale, the User Experience Questionnaire, and the Positive and Negative Affect Schedule. The participants' subjective ratings also indicate the game version was more rewarding and engaging.
In another work, Greipl et al. 
\cite{greipl2021brain} used fMRI and MRI to compare the effects of game-based and non-GBL on the brain. 42 participants played a number line estimation task while their brain activity was measured. The results showed that GBL led to more robust activation in ventral tegmental \& substantia nigra region areas associated with reward and in the amygdala \& anterior insula, indicating emotional processing, which suggests that GBL may enhance learning through rewards. The subjective analysis showed that the game-based version was rated more attractive, novel, and stimulating than the non-game-based version. However, there were no significant differences in the number of correct answers or time taken between the game-based and non-game-based versions of the task.

These results suggest that utilizing Game-Based Learning has the potential to enhance neural activity and elevate engagement levels. These studies have demonstrated enhanced activation in distinct cerebral areas linked to rewards, emotional processing, and cognitive involvement.  In general, the aforementioned findings emphasize the favorable influence of GBL on cerebral activity and involvement. However, some studies struggled to highlight a statistically significant effect for fNIRS.

\subsection{Effectiveness of Serious Game in STEM+C Educational Contexts}
Hsu and Lin \cite{hsu2016impact} compared a Web Digital GBL System in the experimental group with an online VBL system in the control group on computer game programming education. The experimental group exhibited superior learning performance and motivation compared to the control group. A separate investigation conducted by Mohsen 
\cite{ali2016use} explored the effects of virtual surgical simulation on educational achievements. The findings revealed that individuals in the experimental group exhibited better results in language comprehension and vocabulary recognition assessments than those in the control group. Chen et al.
\cite{chen2021learning} conducted a comparative analysis of various degrees of technological engagement in the context of learning. Their findings indicated that using a simulation video game resulted in more substantial advancements in learning outcomes compared to both video-based instruction and traditional instructional methods. Gordillo et al. 
\cite{gordillo2022comparing} studied the efficacy of GBL in software engineering. The researchers discovered that the video games developed by teachers exhibited greater knowledge acquisition and motivation compared to VBL approaches. Tanimoto and Inie
\cite{tanimoto2023creativity} used an online game called ``The Creativity Game" as a pedagogical tool to facilitate instruction and enhance understanding of the theoretical dimensions of creativity. The experiment shows that the game facilitated the acquisition of knowledge related to exploration, value, novelty, limitations, and transformation by the players in a fun and interactive way by participating in many sessions, allowing for the evaluation of uniqueness in relation to the individual. Additionally, the game incorporated simulated critics who provide commentary on the actions taken by the player. 

These studies support the efficacy of game-based and simulation-based learning approaches.
These approaches have demonstrated enhanced learning performance, increased motivation, enhanced language comprehension, and improved knowledge acquisition compared to conventional instructional methods and VBL approaches.

There is a growing body of literature comparing GBL and VBL methods, yet a lack of research focuses on neural correlates associated with these differences.

\section{Materials and Methods} \label{Materials and Methods}

This study aims to explore the efficacy of using serious games for educational purposes with a specific focus on enhancing learning in STEM (Science, Technology, Engineering, and Mathematics) and computer science subjects, particularly within the context of graph structure learning. The primary objective is to assess and compare engagement levels of learners when exposed to serious game-based learning and video-based learning. The evaluation is based on the hemodynamic response and knowledge gain derived from pre- and post-test scores. To achieve this goal, we used functional near-infrared spectroscopy (fNIRS) to measure changes in oxygenation and deoxygenation in participants' prefrontal cortex (PFC). Pre- and post-tests were administered to assess the participants' knowledge and comprehension before and after implementing the learning methods. Furthermore, a survey was conducted to evaluate the usability and task load of the study. To demonstrate the comparison, we divided our participants into two groups: the game group, who played a quiz-based serious game, and the video group, who viewed a recorded video on the same topic.

\subsection{Participants}\label{label:participants}
The study was approved by the University of Delaware's Institutional Review Board (Protocol \#$1982569-1
$). Participants were adults over the age of 18, proficient in the English language, not sensitive to an alcohol rub, and with limited or no familiarity with fundamental graph terminologies. Individuals who did not meet these criteria were excluded during the pre-screening phase. Subsequently, participants received a clear explanation of the terms and conditions of the experiment, as well as the experimental procedures. They were provided informed consent by signing a formal document. Demographic information, including age, courses, year of study, gender, and prior knowledge of graph theory, was collected.

Twelve graduate students (4 female, age = 27.83\,$\pm$\,2.99) took part in this experiment. None of the participants were excluded. There were six students (1 female) in the game group, and in the video group, there were six students (3 females).

\subsection{Apparatus}
\begin{figure}
\centering
\includegraphics[width=0.7\linewidth]{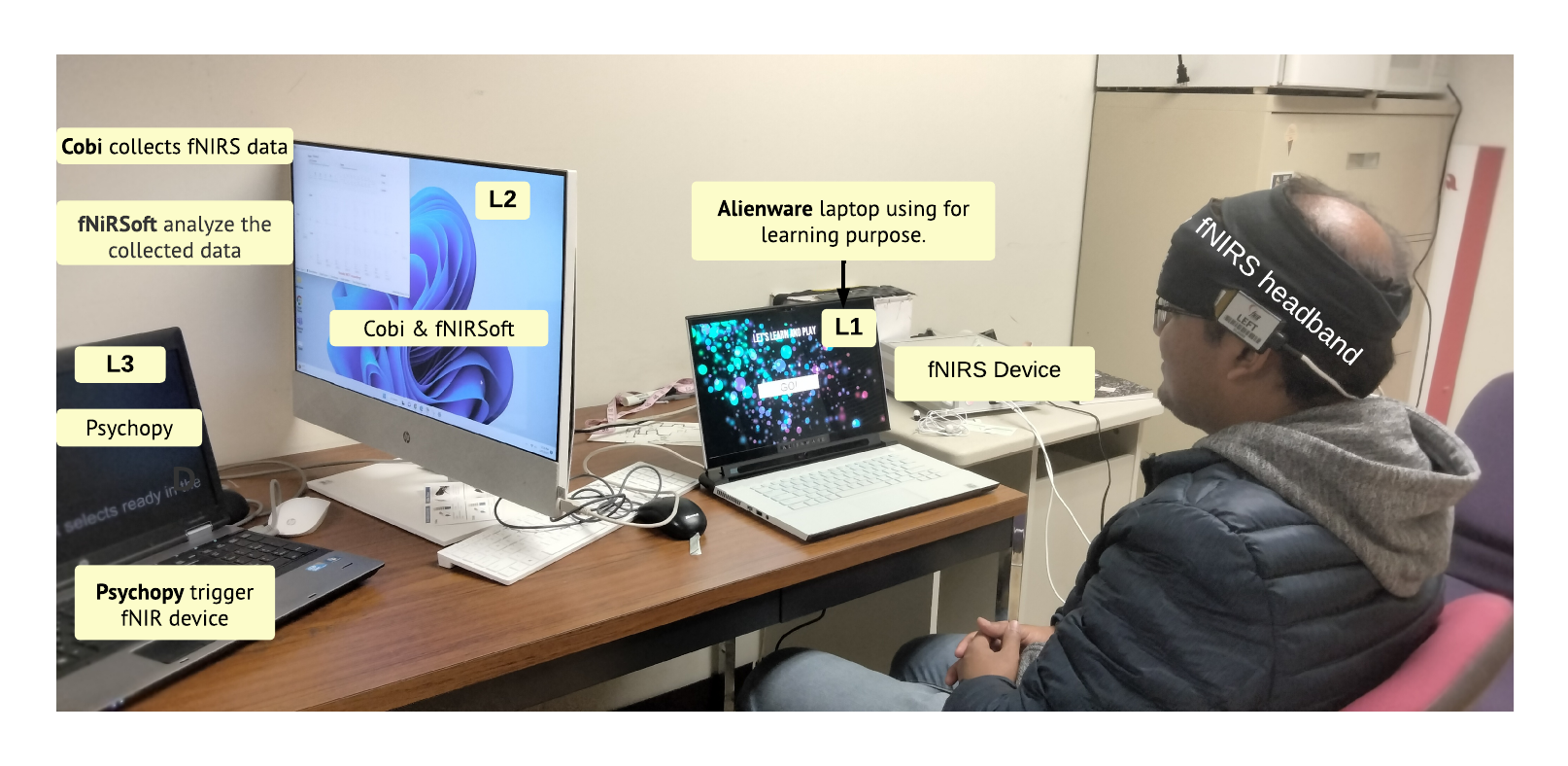}
 
  \caption{Experimental setup for fNIRS data collection: Laptop 1 (L1) is used for learning. Laptop 2 (L2) is connected to the fNIRS device and runs COBI software for data collection. Laptop 3 (L3) sends triggers to mark separate task blocks.}
 
  \label{fig:HwSw}
\end{figure}

We used a combination of hardware and software to carry out the experiment. The hardware and setups are shown in Figure~\ref{fig:HwSw}. 
Three computers were used in the experiment:
\begin{itemize}

\item \textbf{L1 } An Alienware laptop was used to run the learning materials for both the game and video content. 

\item \textbf{L2} A desktop computer was equipped with an Intel (R) Core (TM) i7-10700T CPU operating at a frequency of 2.00 GHz. L2 was connected to an fNIRS device for data acquisition. L2 also ran two key software programs: Cognitive Optical Brain Imaging Studio Software (COBI), responsible for collecting fNIRS signals, and fNIRSoft Software  (Version 4.9), which analyzed the brain signal data by representing the average activation across all blocks for each condition.

\item \textbf{L3} Another laptop with an Intel (R) Core (TM) i5 CPU M 460 @ 2.53 GHz and a serial port. L3 was connected to L2 via serial port. L3 played a key role in the experiment by running a custom PsychoPy \cite{peirce2019psychopy2} code for stimulus presentation and triggering the fNIRS device. PsychoPy sent markers to each block so brain signal data could be segmented properly as separate blocks. The markers included concept definition, rest, quiz, and feedback to L2.

\end{itemize}
The game has been developed using Unity, and the video is prepared using a PowerPoint slide with animation and audio. Python was the selected tool for data analysis and processing. In addition, we used Qualtrics to gather demographic data, conduct pre-and post-tests, and administer user feedback questionnaires.

\subsection{Procedure}
\begin{figure}
\centering
\includegraphics[width=\linewidth]{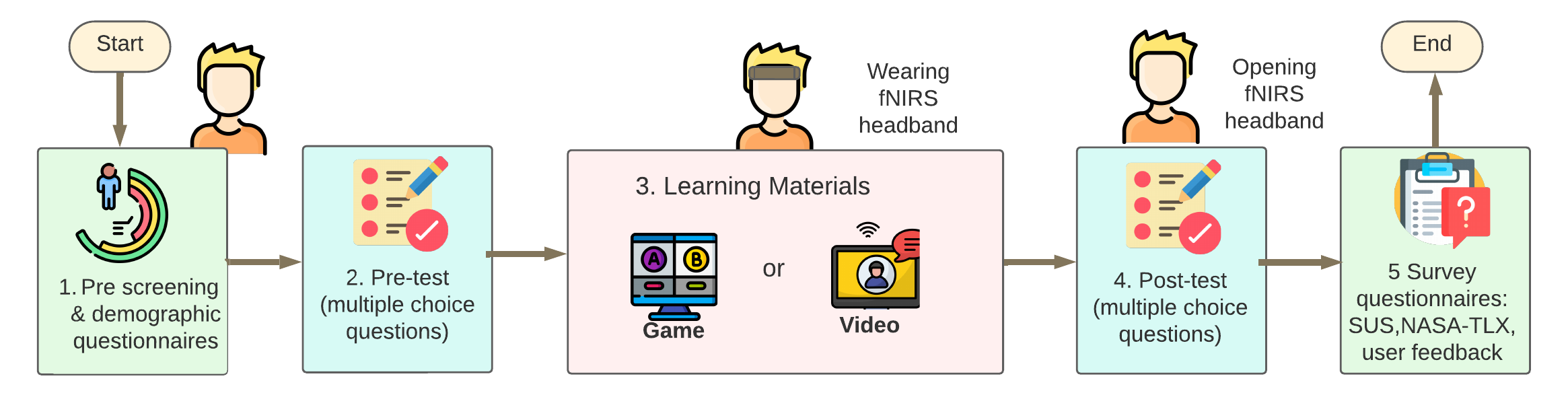}
\caption{Experimental phases: (\textbf{phase 1}) pre-screening and demographic questionnaires; (\textbf{phase 2}) pre-test(\textbf{phase 3})learning material: either game or video wearing fNIRS headband; (\textbf{phase 4}) post-test containing multiple-choice; \textbf{(phase 5)} appear in survey questionnaires}
\label{fig:study_design}
\end{figure}

The experiment was divided into multiple phases, as illustrated in Figure~\ref{fig:study_design}. Initially, the participant did not wear an fNIRS headband.
After consent, participants completed pre-screening questionnaires based on the inclusion and exclusion criteria (section \ref{label:participants}) \textbf{(phase 1)}. Subsequently, participants attended a ten-question pre-test (section\ref{pretest}) and did not get answers or feedback (\textbf{phase 2}).
The next stage involved implementing the instructional approach (section \ref{label_learningMethod}) (\textbf{phase 3}). Two distinct groups were involved: the game-based group played an educational game, and the video-based group watched the video to learn the basic terminologies of graphs recorded on a similar topic. An fNIRS headband was placed on the forehead at the beginning of the learning period and was removed upon session completion.
Following the designated learning period, participants completed the post-test (section \ref{label_posttest}) (\textbf{phase 4}). Participants were not informed about the content of the post-test questions in advance to ensure unbiased responses.
Finally, participants were presented with survey questions about their learning experiences \textbf{(phase 5)}.

\subsection{Design}
This study had a between-subjects design with two conditions: participants were randomly selected to engage in gameplay (N=6) or video content (N=6). 

\subsubsection{Pre-test} \label{pretest}
\hfill\\
During the initial assessment phase, a Qualtrics form was generated, consisting of ten multiple-choice questions (one point each) with five options for each question. 
Participants did not get any feedback about their performance. The test assessed the participants' understanding of the subject matter.
\subsubsection{Learning Methods}\label{label_learningMethod}
\begin{figure*}
\centering
\includegraphics[width=\linewidth]{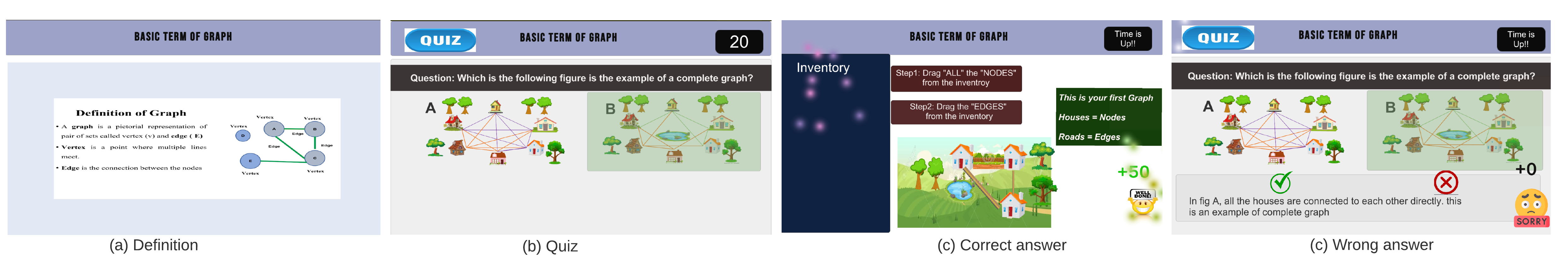}
  \caption{Snapshots of Learning Task Phases and Feedback: Snapshots showing different phases of a learning task: (a) a definition, (b) a quiz question, (c) positive feedback with a score of 50, and (d) negative feedback with a score of zero. 
  }
  \label{fig:game}
  \vspace{-0.3cm}
\end{figure*}

\hfill\\
This study used two instructional conditions, a game, and a video, to teach basic graph terminologies. The selected topics included the definition of a graph, the complete graph, the determination of the number of edges in a complete graph, the concept of a loop, the distinction between directed and undirected graphs, the degree of a graph, and the final concept of in-degree and out-degree in a graph. Figure~\ref{fig:game} shows some components of the definition, quiz, and feedback interface. Both conditions presented questions and contents using the same figures and options, set against a similar environment for consistency of the game and video conditions.

\subsubsection{\textbf{Game-based Learning Interface}}
Players followed the instructions to engage with the game, employing actions such as playing with their graph through drag-and-drop functionality or clicking the answer. After each definition, participants were asked a quiz based on the corresponding term.
The game comprises a total of seven trials, with each trial having a duration of 90 seconds. Each trial has different parts: a definition period lasting 30 seconds and a rest period of 10 seconds. Subsequently, there is a quiz/drag-and-drop period of 10 seconds, followed by a feedback period of 10 seconds. Finally, another rest period concludes each trial. The game lasts approximately 10.5 minutes. The quiz also has game elements such as a timer, sound effects, confetti, clicking, and dragging and dropping options. 

\subsubsection{\textbf{Video-based Learning Interface}}
The instructional video, lasting 10.5 minutes, covers similar topics and features the exact same pictures, time frames for definitions, quizzes, and rest periods as the game content.
The video group verbally responded to each question without engaging with the screen. 
In contrast to the game group, the video screen did not display a timer. The participants could view the correct answers to quiz questions presented within the recorded video.
In contrast to the game group, individuals in this particular group did not receive a numerical score or any form of evaluative feedback, did not have any interactions with the contents, and verbally responded.

\subsubsection{Post-test and Survey Questions} \label{label_posttest}
\hfill\\
After completing the learning methods, a post-test was carried out using a Qualtrics form. The post-test consisted of ten multiple-choice questions (1 point each) 
The order of the questions was randomized for each participant and no feedback was provided to the participants.

After completing the post-test, we gave questionnaires to our participants to assess the learning technique comprehensively described in the section \ref{sec:userExp}.

\subsection{Measures}\label{sec:measure} 
\subsubsection{\textbf{Brain Signal Measure (Quantitative)}}
\begin{figure}
\centering
\includegraphics[width=0.7\linewidth]{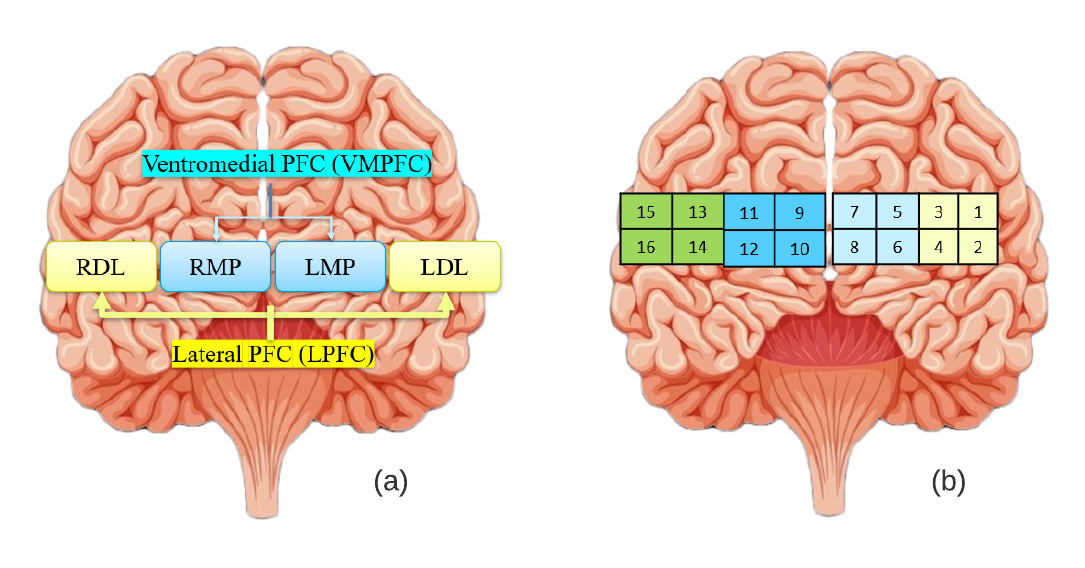}
\caption{(a) Prefrontal Cortex (PFC): LDL, LMP, RMP, and RDL (b) PFC with corresponding fNIRS region \cite{getchell2023understanding}}
\label{fig:brain}
\end{figure}

As noted before, our study used fNIRS to measure the hemodynamic response ($\Delta HbO$ and $\Delta$HbR). 
The prefrontal cortex comprises four distinct regions, as depicted in Figure~\ref{fig:brain} (a). The regions of interest in this study include the Left Dorsolateral Prefrontal Cortex (LDL), Left AnteroMedial Prefrontal Cortex (LMP), Right AnteroMedial Prefrontal Cortex (RMP), and Right Dorsolateral Prefrontal Cortex (RDL) \cite{getchell2023understanding,koiler2022impact, milla2019does}. The four regions mentioned in the study are categorized into two groups: Ventromedial PFC (VMPFC), which comprises the LMP and RMP, and Lateral PFC (LPFC), which encompasses the LDL and RDL. \cite{getchell2023understanding,koiler2022impact, milla2019does}. 
The four regions of the prefrontal cortex were categorized as Region 1 (LDL, optodes 1-4), Region 2 (LMP, optodes 5-8), Region 3 (RMP, optodes 9-12), and Region 4 (RDL, optodes 13-16), based on the optode position of the fNIRS headband (see Fig. \ref{fig:brain}(b)). The fNIRS technique is utilized to quantify the changes in the concentrations of $\Delta HbO$ and $\Delta HbR$ in the prefrontal cortex of study participants.

Initially, calculating and comparing the changes in $\Delta HbO$ and $\Delta HbR$ were conducted across four distinct regions (LDL, LMP, RMP, and RDL) under both game and video conditions. Next, we examined the activation level of $\Delta HbO$ and $\Delta HbR$ in the LPFC and VMPFC during the learning process. Finally, we determined the pre-frontal cortex's overall blood flow of $\Delta HbO$ and $\Delta HbR$.

\subsubsection*{\textbf{$\Delta HbO$ and $\Delta HbR$ Four Regions of PFC}}
The prefrontal cortex comprises four distinct regions \cite{getchell2023understanding, koiler2022impact}. The calculations of $\Delta HbO$ and $\Delta HbR$ are performed for each region, and the mean and standard deviation are determined for each region's game and video conditions. 

\subsubsection*{\textbf{$\Delta HbO$ and $\Delta HbR$ in the LPFC and VMPFC}}
The ratio of LPFC  and VMPFC was calculated following equation (\ref{eq:RNE}) to compare the activity of neurons in the game and video conditions. 
The region of interest (ROI) refers to either LPFC or VMPFC.
\begin{equation}
\small
\label{eq:RNE}
    \textrm{$\Delta HbO$ or    $\Delta HbR$   ratio} 
       =\frac{ \textrm{Game Group ROI}} {\textrm{Video Group ROI}}
\end{equation}

\subsubsection*{\textbf{$\Delta HbO$ and $\Delta HbR$ in Overall PFC}}

Finally, the total value of $\Delta HbO$ and $\Delta HbR$ within the prefrontal cortex was measured to determine the brain activities in the neuron under different study conditions.
Equation~\ref{eq:averageHbO} to get the mean value of $\Delta HbO$ and $\Delta HbR$ for overall PFC.


\begin{equation}
\label{eq:averageHbO}
    \textrm{$\Delta HbO$ or $\Delta HbR$ in PFC} \\
    =\frac{ \textrm{Sum  ($\Delta HbO$ or $\Delta HbR$)}} {\textrm{4}}
\end{equation}
\subsubsection{\textbf{Knowledge Gain (Quantitative)}}
To find the knowledge gain, we calculated the difference between the pre-and post-test by following the equation (\ref{eq:knowledge_gain}). \cite{becker2000analysis}.
\begin{equation} 
\small
\label{eq:knowledge_gain}
    \textrm{Knowledge gain} = 
    \textrm{pre-test score}- \textrm{pre-test score}
\end{equation}

We also calculated the percentage difference of the pre-and post-test for both game and video conditions using the following equation (\ref{eq:percentagePrePost}):
\begin{equation}
\small
\label{eq:percentagePrePost}
    \textrm{Percentage difference}=
    \frac  {  \textrm{diff{(between two scores)}}}{ \textrm{max(between two scores)}} * 100
\end{equation}
\subsubsection{\textbf{User Experience} }\label{sec:userExp}

We used the System Usability Scale (SUS) with a range between 0 and 100\% (0–50\%: not acceptable, 51–67\%: poor, 68\%: okay, 69–80\%: good, 81–100\%: excellent), a reliable tool for measuring the usability of our learning methods.  \cite{brooke1996sus}.
We used the NASA task load index (NASA TLX), a survey instrument to measure and perform a subjective mental workload assessment to determine the load of a participant while performing the learning methods with six questions to determine an overall workload rating \cite{hart1988development}.
Participants were also asked to rate their experience using a five-point Likert scale ranging from 1 (low) to 5 (high) based on five questions concerning motivation, interaction, fun, engagement, and brainstorming.


\begin{itemize}
\item Motivating: The learning methodology was motivating			
\item Fun: The learning methodology was fun			
\item Interactive: I found this learning method interactive			
\item Engaging: I found this learning method was engaging me with the topic effectively	
\item Brainstorming: The learning method allows me to brainstorm while learning the terminologies	
\end{itemize}

\section{Results and Discussion} \label{Result}
This section discusses the findings derived from analyzing brain signals, questionnaires, and knowledge gain ratings.  

\subsection{Brain Signal Analysis}
The study showed different patterns of cerebral activity in the group that engaged with the game and the group that watched the video. 
Using fNIRSoft software, we obtained an Excel file containing time series data for oxygenated hemoglobin ($\Delta HbO$) and deoxygenated hemoglobin ($\Delta HbR$) values from each optode. We analyzed the results based on four regions of PFC. We calculated $\Delta HbO$ and $\Delta HbR$ for each of the four regions (LDL, LMP, RMP, and RDL) under both game and video conditions.

\subsubsection{$\Delta HbO$ and $\Delta HbR$ Values in Four Regions of the Brain}
The findings show that the game condition has higher $\Delta HbO$ values across four regions, indicating higher brain activities, as shown in Table \ref{tab:four}. Conversely, the video condition demonstrates higher $\Delta HbR$ values in LMP, RMP, and RDL.
\begin{table}[h]
\caption{Summary of descriptive results of $\Delta HbO$ and $\Delta HbR$ for four regions 1: LDL, 2: LMP, 3: RMP, 4: RDL}
\label{tab:four}
\centering
\begin{tabular}{l|ll|ll}\hline
                 & \multicolumn{2}{c|}{\textbf{$\Delta HbO$}}                              & \multicolumn{2}{c}{\textbf{$\Delta HbR$}}                             \\\hline 
\textbf{Region} & \multicolumn{1}{c}{\textbf{Game}} & \multicolumn{1}{c|}{\textbf{Video}} & \multicolumn{1}{c}{\textbf{Game}} & \multicolumn{1}{c}{\textbf{Video}} \\\hline 
LDL              & 0.33 (0.61)            & 0.21 (0.67)              & 0.25 (0.57)             & 0.1 (0.2)                \\
LMP              & 0.16 (0.3)             & -1.21 (2.59)             & 0.16 (0.32)            & 0.42 (1.07)          \\
RMP              & 0.14 (0.3)            & -1.21 (2.87)            & 0.09 (0.27)          & 0.18 (0.56)             \\
RDL              & 0.23 (0.58)             & 0.03 (0.25)             & 0.17 (0.29)            & 0.36 (0.71)   \\\hline        
\end{tabular}\\
 
\textit{All entities are in the format: mean value (standard deviation).}
\end{table}

\subsubsection{$\Delta HbO$ and $\Delta HbR$ Values in the LPFC and VMPFC}
The LPFC shows increased oxygen consumption with higher $\Delta HbO$, which means higher brain activities during the learning period in both game and video conditions, as shown in Table \ref{tab:LPFC VMPFC}. Interestingly, in the game condition, $\Delta HbO$ (M=0.28) in the LPFC is 2.33 times higher compared to the video condition (M=0.12). However, the $\Delta HbR$ value in the video condition (M=0.23) is slightly higher than in the game condition (M=0.21).


\begin{table}[h]
\caption{Summary of descriptive results of $\Delta HbO$ and $\Delta HbR$ value for LPFC and VMPFC}
\label{tab:LPFC VMPFC}
\centering
\begin{tabular}{l|cc|cc} \hline
      & \multicolumn{2}{c|}{\textbf{$\Delta HbO$}}                         & \multicolumn{2}{c}{\textbf{$\Delta HbR$}}               \\\hline
   \textbf{Condition}   & \textbf{LPFC}                  & \textbf{VMPFC}                   & \textbf{LPFC}                   & \textbf{VMPFC}                  \\\hline
\textbf{Game}  & 0.28 (0.56)  & 0.15 (0.3)   & 0.21 (0.38)  & 0.12 (0.25)  \\
\textbf{Video} & 0.12 (0.4)  & -1.21 (2.72)  & 0.23 (0.43)  & 0.3 (0.81)  \\ \hline
\end{tabular}\\
 
\textit{All entities are in the format: mean value (standard deviation).}
\end{table}

\subsubsection{$\Delta HbO$ and $\Delta HbR$ Values in the PFC}
\begin{figure} [h]

  \centering
  \includegraphics[width=0.68\linewidth]{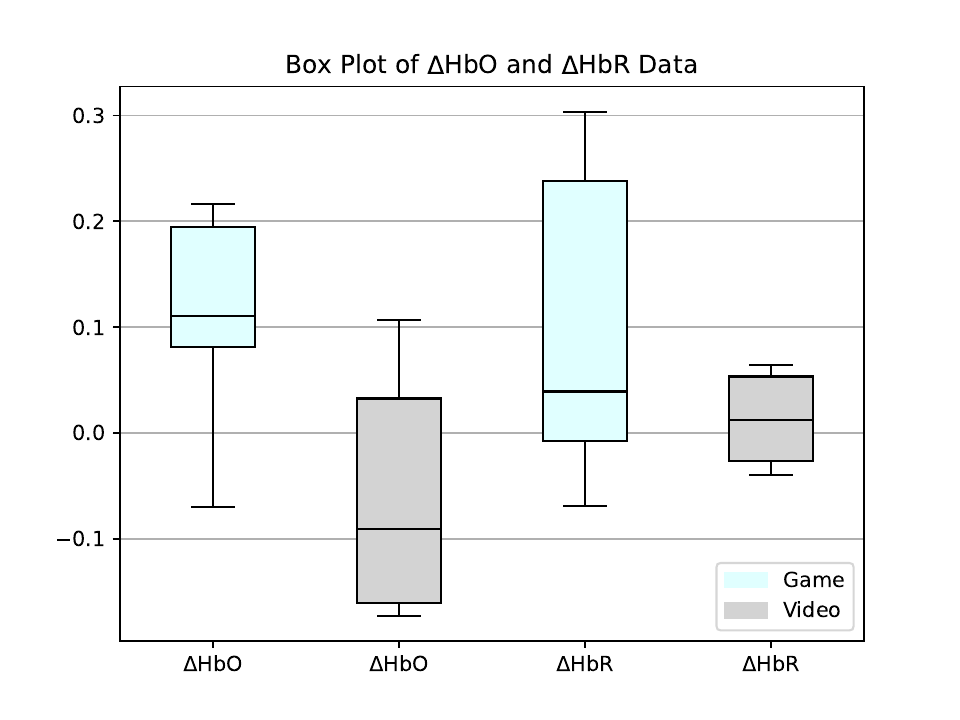}

\caption{Comparison of mean $\Delta HbO$ and $\Delta HbR$ for Game and Video condition. Higher values represent more neural activities in the brain}  \label{fig:delHbRHbOBox}
\end{figure}

The game condition has higher $\Delta HbO$ (M=0.21) values than the video condition (M=-0.54), indicating more neural activities.
Participants who played the game exhibited a 357.14\% greater increase in oxygen utilization than those who watched the video. However, $\Delta HbR$ is almost the same in the video (M=0.26) and game (M=0.16) conditions, as shown in Figure~\ref{fig:delHbRHbOBox}. 
There is no significant difference between the game and video conditions in $\Delta HbO$ and $\Delta HbR$ values.
Overall, our data from six participants in each game group were shown to have a wide standard deviation, which is a known challenge in the research community.

\vspace{-0.2cm}
\subsection{Knowledge Gain}

\begin{figure}

\centering
\subfloat[Knowledge gain score difference]{\includegraphics[width=0.2\textwidth]{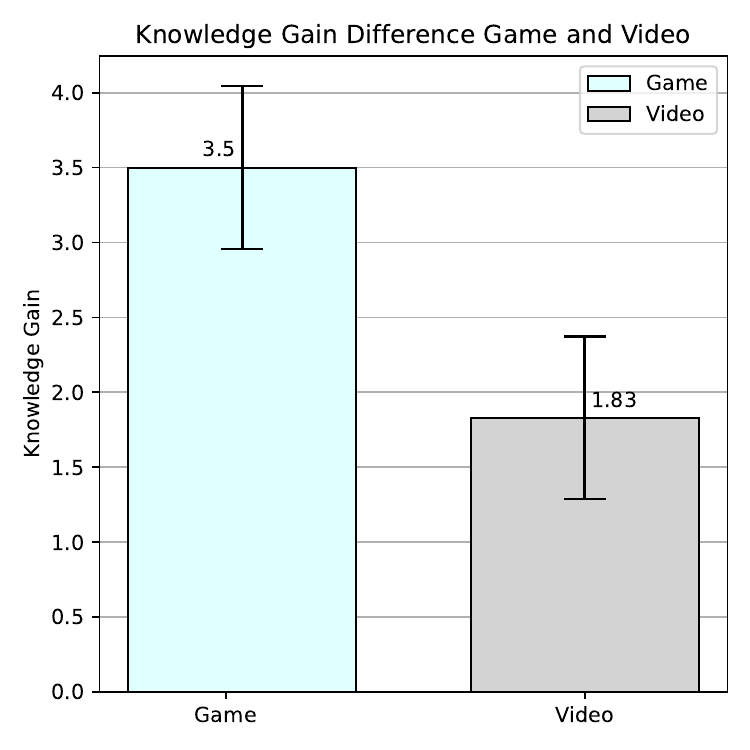}} \hskip1ex
\subfloat[The difference of the pre-and post-test Score]{\includegraphics[width=0.2\textwidth]{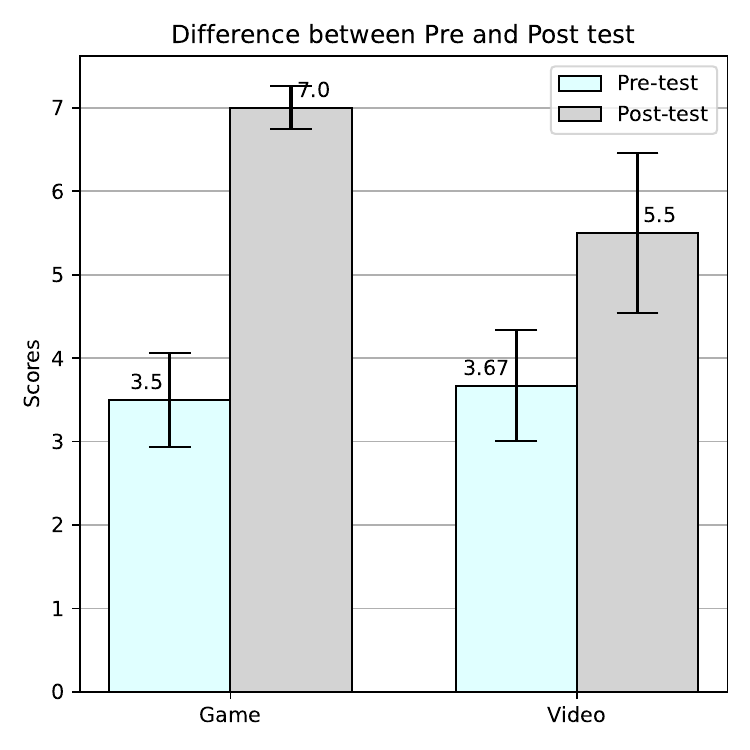}}
\caption{(a) Score Differences in Game vs Video Conditions (b) Pre-Post Test Score Differences}
\label{fig:scoreDiff}
\end{figure}

Figure~\ref{fig:scoreDiff} (a) shows that the game condition (M=3.5 $\pm$ 1.52) exhibits higher knowledge gain compared to the video condition (M=1.83 $\pm$ 1.33). The game group's knowledge gain is 47.74\% higher than that of the video group, as calculated in Equation~\ref{eq:percentagePrePost}. Statistical analysis showed no statistical difference between the game and video conditions (t=2.02,  p=0.07) where p-value is close to the critical threshold of alpha =0.05.

Figure~\ref{fig:scoreDiff} (b) illustrates the percentage difference between the pre-and post-tests. The game group shows a higher percentage difference (50\%) compared to the video group (33.33\%), with a 16.67\% difference between the two.

\subsection{User Experience}
\subsubsection{Usability}
\begin{figure}[h]
\centering
\includegraphics[width=0.99\linewidth]{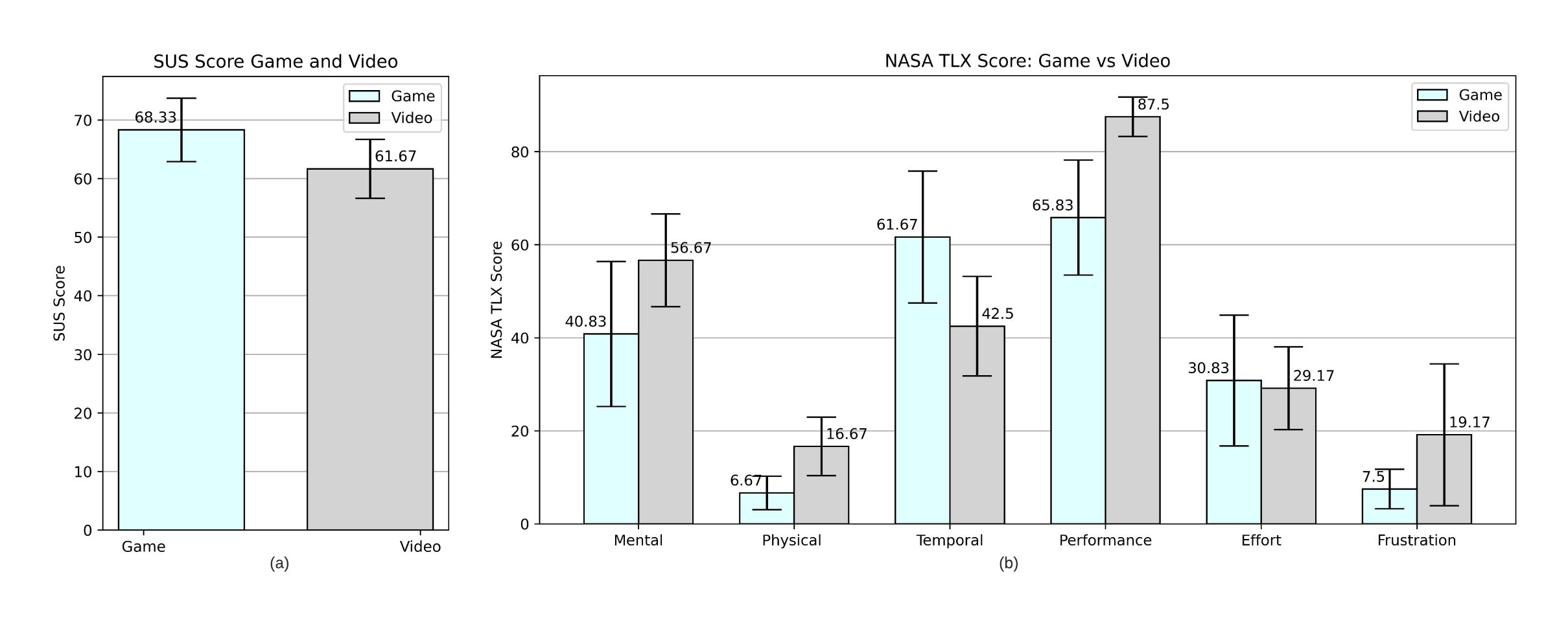}
 \caption{(a) System usability scale (SUS) (b)Task load score (NASA-TLX)}
  \label{fig:NASA_score}

\end{figure}

The SUS result shows that the average score of the game condition (M = 68.33 $\pm$ 13.29) is relatively higher than the video condition (M = 61.67$\pm$ 12.32), as shown in Figure~\ref{fig:NASA_score} (a). 
 ANOVA analysis on SUS score did not show statistically significant differences.


\subsubsection{Task Load}
The overall workload scores show an average of M = 25 for the game condition and M = 51.67 for the video condition (lower scores indicate lower workload). Descriptive results show that the video condition exhibits higher average scores for mental demand, physical demand, performance, and frustration, while temporal demand and effort are slightly lower than in the game condition, as shown in Figure~\ref{fig:NASA_score} (right). Due to participant variability and small sample size, no statistically significant difference was observed between the game and video conditions for NASA-TLX.

\subsubsection{Engagement and Motivation}
\begin{figure} [h]
\centering
  \includegraphics[width=0.99\linewidth]{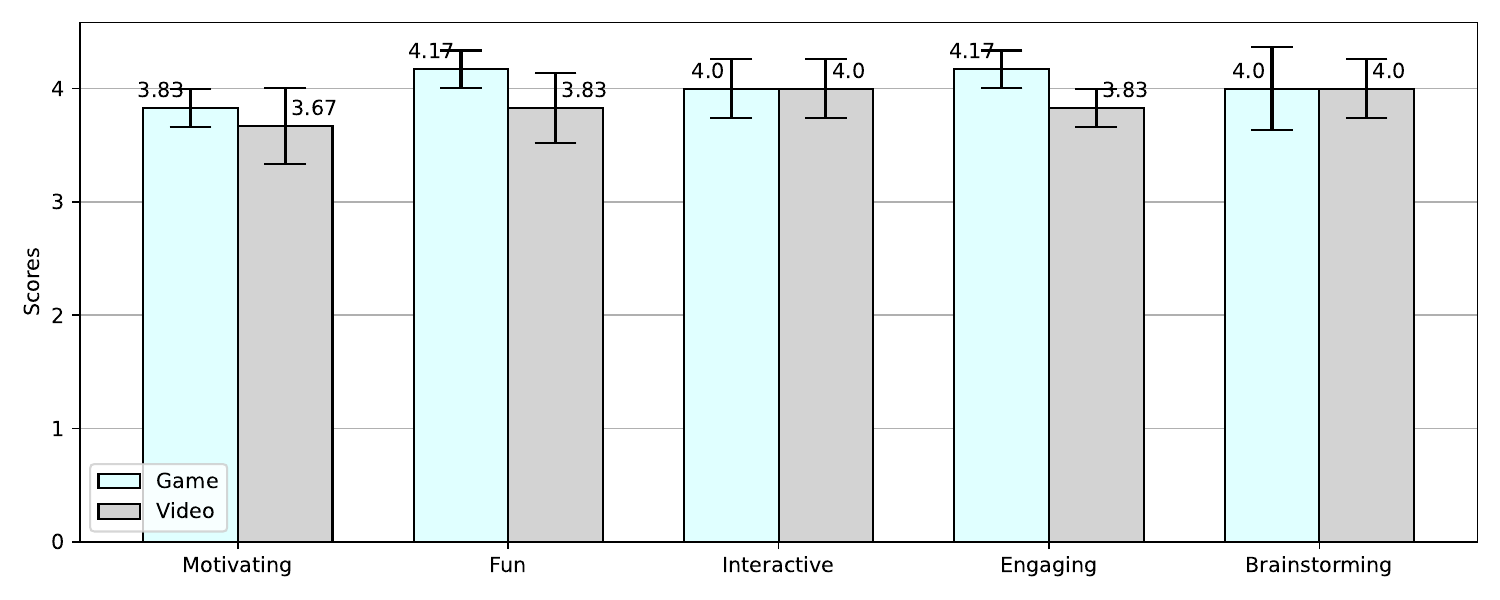}
  \caption{User Experience in Engagement and Motivation}
  \label{fig:fun}
\end{figure}

Figure~\ref{fig:fun} shows the average results for user experience ratings for game and video conditions. Interestingly, we did not find any significant effects of motivation, engagement, etc., between the game and video conditions in terms of motivation, fun, engagement, interaction, and brainstorming. 
Participants reported similar levels of engagement in both the game and video conditions. This can be due to video conditions using the majority of the game elements, thus being as engaging and interactive as game conditions. 

\section{Discussion}\label{Discussion}


In this section, we discuss our findings based on the research questions:

For \textbf{RQ1}, the goal was to identify the differences in neural activities between game and video learning. 
The prefrontal cortex oxygenation pattern analysis showed no significant effect of the change of $\Delta HbO$ and $\Delta HbR$ between the game and video groups. However, the game group exhibited a higher  $\Delta HbO$ mean value compared to the video group. 
The observations also found substantial dispersion variability among participants. This was a between-subjects study, and it remained a challenge to gain statistically significant results in our preliminary study, but we are aiming to resolve some of the challenges in the following experiments.
The $\Delta HbR$ concentration change during the learning period shows a small difference between the game group
and the video group
. This suggests that participants exhibit slightly higher neural activity during gameplay compared to video watching.
Overall, the findings suggest that there may be increased oxygenation and neural activity during gameplay compared to video watching.
Comparing the $\Delta HbO$ levels between the LPFC and the VMPFC, LPFC consistently exhibited higher values, indicating higher neural activity. 

For \textbf{RQ2}, the aim was to capture the differences regarding the subjective results on usability, task load, and knowledge gain between game and video learning. The utilization of gaming interfaces, which include timer mechanisms, drag-and-drop functionality, and click-based interactions, along with immediate feedback in the form of sounds, visual elements such as emojis, scoring systems, and celebratory animations like confetti, within the game, may have contributed to an increased oxygenation flow. This is also evidenced by the knowledge gained from the results, with the game-based approach showing a 47.74\% higher increase in knowledge compared to the video group. Additionally, user experience questionnaires indicate that the game-based approach is perceived as more user-friendly and enjoyable compared to the video-based method. 
Interestingly, when participants were asked to evaluate the interactivity of both interfaces, the results revealed an equal score for both the game and video interfaces. 
This could be explained by the fact that some participants in the video group have found responding with their voices to an interactive element, despite differences in how they engaged compared to the game interface.
\section{Conclusion}
In this work, we introduce and compare two graph structure education modules of game and video by examining participants' prefrontal cortex oxygenation patterns. In a pilot user study with twelve participants, we used the fNIRS device to measure the changes in the prefrontal cortex's oxygenated hemoglobin ($\Delta HbO$) and deoxygenated hemoglobin ($\Delta HbR$) levels. This was a preliminary study that was presented in our study. We noted that the mean levels of oxygenated hemoglobin ($\Delta HbO$) were higher in the GBL group, suggesting the potential enhanced cognitive involvement. Additionally, the lateral prefrontal cortex (LPFC) had greater hemodynamic activity during the learning period. Moreover, knowledge gain analysis showed an increase in mean score in the game group compared to the video group. Although we did not observe statistically significant changes due to participant variability and sample size, this preliminary work contributes to understanding how game- and video-based learning impact cognitive processes, providing insights for enhanced instructional design and educational game development. 

Future research should explore the potential benefits of specific game design elements and their impact on educational outcomes. Expanding the sample size and enhancing the gaming interface can help mitigate constraints present in existing research.
In subsequent studies, we intend to improve the educational modules as explained before and eventually test them with a group of high school students.
Furthermore, we plan to integrate this game into Augmented Reality (AR) and Virtual Reality (VR) platforms, allowing participants to fully immerse themselves in a realistic environment and enhancing their engagement. Additionally, we aim to qualitatively evaluate the system by monitoring users' facial expressions and tracking their gaze as a supplement to self-reported surveys and assessments.

\section*{Acknowledgment}
We wish to express our gratitude to the study participants and lab members. We also thank the support from the National Science Foundation(\#$2222661-2222663$ and \#$2321274$). Any opinions, findings, and conclusions expressed in this material are those of the authors and do not necessarily reflect the view of sponsors.
\newpage

\balance
\bibliographystyle{IEEEtran}
\bibliography{IEEEabrv,bibfile}

\end{document}